\documentclass[aps,prl,twocolumn,
amsmath,amssymb,superscriptaddress]{revtex4-2}
\usepackage[utf8]{inputenc}
\usepackage{amsmath}
\usepackage{ulem}
\usepackage{graphicx}
\usepackage{float}
\usepackage{hyperref}
\usepackage{lipsum}
\usepackage{listings}
\usepackage{listingsutf8}
\usepackage{xcolor}
\usepackage{amssymb}
\usepackage{textcomp}
\usepackage{units}
\usepackage{bbm}
\usepackage{enumitem}  
\usepackage{url}
\usepackage[english]{isodate}
\usepackage{bm}
\usepackage{mdframed}
\usepackage[capitalise]{cleveref}
\usepackage{enumerate}
\usepackage{dsfont}
\usepackage{dcolumn}
\usepackage{physics}
\usepackage{color}
\usepackage{placeins}
\usepackage{slashed}
\usepackage{amssymb}
\usepackage{hyperref}
\usepackage[capitalise]{cleveref}
\usepackage[caption=false]{subfig}
\usepackage{braket}
\usepackage{xspace}

\newcommand{\cL}{\mathcal{L}}

\newcommand*{\NNLO}{N2LO\xspace}

\newcommand*{\Mpi}{\ensuremath{m_\pi}}

\newcommand*{\Mhi}{M_\text{hi}}

\newcommand{\prob}{p}

\begin{document}

\title{Quantifying the breakdown scale of pionless effective field theory}
\author{Andreas Ekström}
\affiliation{Department of Physics, Chalmers University of Technology, SE-412 96, Göteborg, Sweden}
\author{Lucas Platter}
\affiliation{Department of Physics and Astronomy, University of Tennessee, Knoxville, Tennessee 37996, USA}
\affiliation{Physics Division, Oak Ridge National Laboratory, Oak Ridge, Tennessee 37831, USA}

\date{\today}
\noindent
\begin{abstract} 
We use Bayesian statistics to infer the breakdown scale of pionless effective field theory in its standard power counting and with renormalization of observables carried out using the power-divergence subtraction scheme and cutoff regularization. We condition our inference on predictions of the total neutron-proton scattering cross section up next-to-next-to leading order. We quantify a median breakdown scale of approximately 1.4~$\Mpi$. The 68\% degree of belief interval is $[0.96,1.69]\Mpi$. This result confirms the canonical expectation that the pion mass is a relevant scale in low-energy nuclear physics.

\end{abstract}

\maketitle

\paragraph*{\bf Introduction --} 
Effective field theories (EFTs)~\cite{Georgi:1993mps} have emerged as important tools in nearly all areas of physics. They facilitate precise and systematic descriptions of observables without requiring a complete understanding of an underlying theory by focusing on the relevant degrees of freedom. A requirement for the applicability of an EFT is the existence of a sufficiently large separation of scales inherent to the system under study, as only then does the ratio of these scales provide a useful expansion parameter.

One type of EFT, the short-range EFT~\cite{vanKolck:1998bw,Kaplan:1998we,Kaplan:1998tg}, has found wide application in particle, nuclear, and atomic physics. This non-relativistic EFT is built solely from contact interactions and is applicable when the two-body scattering length, $a$, is much larger than the range, $R$, of the interaction. In atomic physics, short-range EFT has been used to analyze three-body recombination in ultra-cold atomic gases and to relate its loss features to the Efimov effect~\cite{Braaten:2004rn}. In nuclear physics, it has been applied to describe a wide range of low-energy phenomena in the positive- and negative-energy spectra of light-mass nuclear systems~\cite{Hammer:2019poc}. Moreover, this type of EFT has enabled the calculation of electroweak processes in the two- and three-nucleon systems and served as a framework for describing electroweak processes involving halo nuclei, consisting of a few nucleons weakly bound to a tightly bound core~\cite{Hammer:2022lhx}. In particle physics, contact EFT is a powerful tool to understand the properties of weakly bound mesonic molecules like the X(3872)~\cite{Braaten:2003he}.

Given that EFTs offer a methodical order-by-order approach, they were heralded as providing reliable uncertainty estimates. To quantify these uncertainties---both from truncating the EFT expansion~\cite{Furnstahl:2015rha} and in parameter estimation~\cite{Svensson:2021lzs}---Bayesian methods were developed. Recently, it has been recognized that Bayesian approaches can be used to quantify the breakdown scale, $\Mhi$, of an EFT~\cite{Melendez:2019izc}. The breakdown scale crucially determines the momentum-scale for which the EFT is expected to fail, though it is not always straightforward to quantify. Indeed, until we have quantitative knowledge about the properties of the underlying theory, low-energy quantum chromodynamics in the case of nucleons, $\Mhi$ remains an inferred quantity rather than a precisely defined one. In the case of pionless EFT (the short-range EFT for nucleons), the canonical expectation for the breakdown scale is momenta corresponding to the pion mass, $\Mpi \approx 138$ MeV. This is because pion exchange---the longest-range nuclear interaction, as described by Yukawa~\cite{Yukawa:1935xg}---is omitted from this EFT.

In this work, we use order-by-order predictions of the the total neutron-proton scattering cross section to quantify $\Mhi$ and thereby also critically test the fundamental underpinnings of pionless EFT~\footnote{During the preparation of this manuscript, we became aware of Ref.~\cite{Bub:2024gyz}, which investigates the breakdown scale of a theory constructed solely from contact interactions in the context of nucleon-nucleon scattering. That work follows a non-standard ordering scheme for subleading interactions, which differs from the approach typically used in pionless EFT. Consequently, its results are not directly comparable to our findings.}. This work highlights the strength of combining Bayesian methods with an order-by-order renormalizable EFT to leverage its inferential advantages.

\paragraph*{\bf Pionless EFT -} This is a field theoretical formulation of effective range theory~\cite{Bethe:1949yr}. It uses a non-relativistic Lagrangian built from contact interactions only 
\begin{equation}
    \label{eq:Lagrangian}
   \cL = N^\dagger ( i\partial_t +\frac{\nabla^2}{2m}) N
   +\sum_{\alpha}\cL_\alpha+ \cL_{sd}~,
\end{equation}
where $N$ denotes the nucleon field and $\alpha = {}^1S_0, {}^3S_1$ are the spin-singlet ($s$) and spin-triplet ($t$) contributions to the $S$-wave two-nucleon scattering channel
\begin{multline}
   \cL_\alpha =  - C^\alpha_0 (N^\dagger P_i^{\alpha} N)^\dagger (N^T P_i^\alpha N)\\
          +\frac{C^\alpha_2}{8}\left[\left(N^TP_i^{\alpha} N \right)^\dagger
    \left(N^T\overleftrightarrow{\nabla}^2 P_i^{\alpha} N\right) +\rm{h.c.}\right]~,
    \label{eq:La}
\end{multline}
where $C^\alpha_0$ and $C^\alpha_2$ are low-energy coupling constants adjusted to the effective range parameters \cite{Chen:1999tn}. At next-to-next-to-leading order (N2LO), an $S$- to $D$-wave operator is formally required to calculate the scattering amplitude, denoted in Eq.~\eqref{eq:Lagrangian} as $\mathcal{L}_{sd}$. However, it does not contribute to the total cross section up to N2LO.

We calibrate the coupling constants of this Lagrangian to reproduce the effective range parameters determined by the Nijmegen partial wave analysis~\cite{deSwart:1995ui,Stoks:1993tb}. To do this, we expand around the pole in the triplet channel
\begin{align}
\label{eq:pole_expansion}
k \cot \delta_t = -\gamma_t +\frac{\rho_t}{2}(k^2 +\gamma_t^2) + \ldots~, 
\end{align}
where $k$ is the relative momentum between the two nucleons, $\gamma = \sqrt{m B_d}$ is the binding momentum associated with the binding energy $B_d$ of the deuteron. In the singlet channel, we expand around the scattering threshold
\begin{align}
\label{eq:threshold_expansion}
k\cot \delta_s = -\frac{1}{a_s}+ \frac{r_s}{2}k^2+\ldots~.
\end{align}
Pionless EFT requires a re-summation of the contact operators at leading order to reproduce the analytic structure of the  $S$-matrix that generates the low-energy bound state (the deuteron) in the triplet channel and a virtual state in the singlet channel. The two-body spin-singlet $t$-matrix $t_s$ containing the effective range parameters is therefore expanded in the small length scale $~ r_s$
\begin{multline}
\label{eq:amplitude_expansion}
t_s(k) = \frac{1}{-\frac{1}{a_s} - i k}\Biggl[
1- \frac{r_s}{2}k^2\frac{1}{-\frac{1}{a_s}- i k} \\
+\left(\frac{r_s}{2}k^2
\frac{1}{-\frac{1}{a_s}-i k}\right)^2+\ldots\Biggr]~.
\end{multline}
A similar expansion is carried out in the triplet channel for the triplet t-matrix $t_t$ where \eqref{eq:pole_expansion} is used. The first term in the expansion in Eq.~\eqref{eq:amplitude_expansion} is reproduced in the pionless EFT by summing all diagrams that contain only vertices arising from the first term in Eq.~\eqref{eq:La} and calibrating the low-energy constant $C_0^{{}^1S_0}$ accordingly. The remaining terms are reproduced by including the subleading operators in Eq.~\eqref{eq:La} in perturbation theory. In this way, the low-energy constants become functions of the parameters in the effective range expansions~\cite{Chen:1999tn}
\begin{align}
\label{eq:parameters}
a_s = -23.714\, {\rm fm}, & \,\,r_s= 2.678\, {\rm fm},\\
\gamma^{-1} = 4.318946\, {\rm fm}, & \,\,\rho_t = 1.765\, {\rm fm}~,
\end{align}
and we calculate the total cross section as
\begin{align}
\sigma(k) = 4\pi\left(
\frac{1}{4}|t_s(k)|^2 + 
\frac{3}{4}|t_t(k)|^2\right)~.
\end{align}
The calculation of the scattering amplitude includes loop diagrams that are power-divergent. Therefore, a regularization scheme has to be employed before renormalization to the physical parameters in Eq.~\eqref{eq:parameters}. The so-called power-divergence subtraction (pds) scheme \cite{Kaplan:1998tg} is a regularization approach that replaces any power divergence with a single power of the renormalization scale $\mu$. After renormalization in the pds scheme, the on-shell two-nucleon $t$-matrix reproduces the effective range expansion up the order of the EFT expansion exactly, {\it i.e.}, it does not exhibit any residual regulator dependence. Alternatively, the loop integrals can employ a hard momentum space cutoff $\Lambda$. When this is done, the amplitude will display residual regulator dependence that is one order higher than the one considered. In this work, we primarily use pds regularization, but we also explore the effects of cutoff regularization on our inferences \footnote{See the supplemental material for more information on regularization and renormalization with a hard momentum space cutoff}.

\paragraph*{\bf Inferring the breakdown scale --}Having established the theoretical framework for pionless EFT, we now turn to the application of Bayesian inference to quantify the breakdown scale $\Mhi$. We express the $n$-th order EFT prediction of the total cross $\sigma$ section, at relative on-shell momentum $k$, as a series expansion, i.e., we formally write
\begin{align}
\label{eq:cs_expansion}
\sigma^{(n)}(k) & = \sigma_\text{ref}(k)
\sum_{i=0}^n c_i(k) [Q(k)]^i~.
\end{align}
The expected systematicity of pionless EFT manifests in the dimensionless ratio $Q(k)=f(k)/\Mhi$ and an expectation of natural values for the EFT expansion coefficients $c_i$. We assume a functional form $f(k)$ that smoothly interpolates over the soft scale $\sim 1/a_t$, where $a_t=5.42$ fm, as
\begin{equation}
f(k;r) = \frac{k^r + (1/a_t)^r}{k^{r-1} + (1/a_t)^{r-1}},\,\, r = 6
\label{eq:soft_scale_function}
\end{equation}
This function is roughly constant for $ka_t<1$ and smoothly matches to a linearly increasing function $k/\Mhi$. In accordance with our expectation of naturally sized expansion coefficients $c_i$, i.e., $c_i \approx \mathcal{O}(1)$, we employ a normally distributed prior density\footnote{The statistics notation $x \sim \cdot$ is shorthand for ``$x$ is distributed as $\ldots$''. The abbreviation iid stands for independent and identically distributed.} 
\begin{align}
c_i | \bar{c}^2 {}&\overset{\text{iid}}{\sim} \mathcal{N}(0,\bar{c}^2) \label{eq:prior1}\\
\bar{c}^2|\text{naturalness}{}&\sim \chi^{-2}(\nu_0=2,\tau_0^2=1),
\label{eq:prior2}
\end{align}
\begin{figure}[t]
  \centerline{\includegraphics[width=0.9\columnwidth]{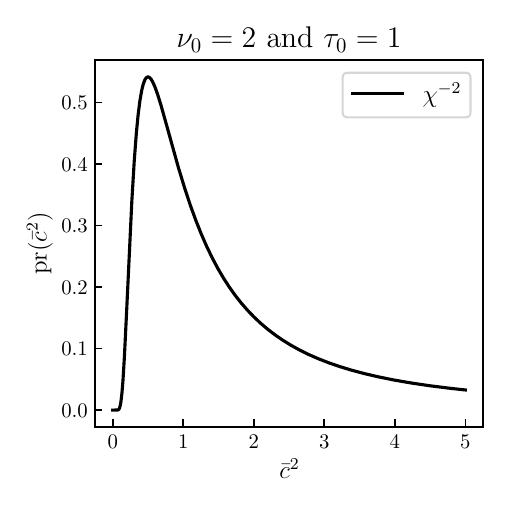}}
  \caption{\label{fig:cbar_prior} Inverse-$\chi^2$ prior for the variance of the normally distributed prior we place on the EFT expansion coefficients $c_i$ defined in Eq.~\eqref{eq:cs_expansion}.}
\end{figure}
with a hyperprior for the variance $\bar{c}^2$ following an inverse $\chi^{2}$ distribution with degrees of freedom $\nu_0=2$ and scale parameter $\tau_0=1$, equivalent  to an inverse gamma distribution $\mathcal{IG}(a_0=1,b_0=1)$, see Fig.~\ref{fig:cbar_prior}. 
For this choice of prior we have $\mathbb{P}(\bar{c}^2 \in [1/3,3])\approx0.67$, i.e. a majority of the probability for the variance remains natural.

All dimensionful factors are collected in a reference scale $\sigma_\text{ref}$ which renders the expansion coefficients $c_i$ dimensionless. Our statistical analysis will be conditioned on leading order (LO,$n=0$), next-to-leading order (NLO $n=1$), and next-to-next-to-leading order (\NNLO, $n=2$) predictions for the total cross section at a finite number of momenta $k$. Information about the breakdown scale flows through the  momentum-dependent order-by-order differences of predictions $\sigma^{(j)}$ and $\sigma^{(j-1)}$ via the corresponding expansion coefficients
\begin{equation}
c_j(k) = \frac{\sigma^{(j)}(k) - \sigma^{(j-1)}(k)}{\sigma_\text{ref}[Q(k)]^j}.
\end{equation}
We utilize the leading-order pionless EFT prediction as the reference scale and therefore have that $c_0=1$. A collection of order-by-order predictions at $K$ different momenta is denoted by a boldface symbol; $\bm{\sigma}$. Up to \NNLO, we can extract $N=2$ informative coefficients $c_1(k)$ and $c_2(k)$ at momenta $(k_1,k_2,\ldots,k_K)$.
\begin{figure}[t]
  \centerline{\includegraphics[width=0.9\columnwidth,angle=0,clip=true]{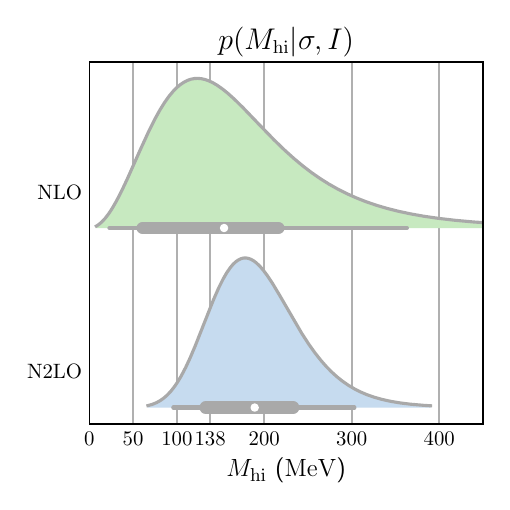}}
  \caption{\label{fig:EFT_breakdown} Posteriors for the breakdown scale $\Mhi$ in pionless EFT. Thick and thin horizontal bars indicate 68\% and 95\% (highest posterior density) DoB intervals, respectively. The 68\% (95\%) DoB intervals are $[61,216]$ MeV ($[23,363]$ MeV) and $[133,233]$ MeV ($[96,303]$ MeV) at NLO and \NNLO, respectively. The median values (white dots) for $\Mhi$ are 154 and 189 MeV at NLO and \NNLO, respectively.}
\end{figure}
Using Bayes' rule, we express the posterior probability for $\Mhi$, given $\bm{\sigma}$ and additional assumptions $I$, specified above and below, as
\begin{equation}
\prob(\Mhi|\bm{\sigma},I) \propto \prob(\bm{\sigma}|\Mhi,I)\prob(\Mhi|I).\label{eq:bayes_rule}
\end{equation}
To avoid overly correlated samples, which would contradict our iid assumption for the expansion coefficients, we computed cross sections at $K=3$ different laboratory scattering energies (10,40,and 70 MeV) corresponding to relative momenta $k=(68.5, 137.0, 181.2)$ MeV~\cite{Svensson:2023twt}. Below we analyze the robustness of our inference with respect to this choice of scattering energies.

The iid assumption enables a factorization of the data likelihood
\begin{equation}
\prob(\bm{\sigma}|\Mhi,I) = \prod_{i=1}^K \prob(\bm{\sigma}_i|\Mhi,I).
\end{equation}
Following~\citet{Melendez:2019izc} we can make a variable transformation and express the data likelihood at each momentum $k_i$ as a joint distribution for the corresponding expansion coefficients. Moreover, having placed a (conjugate) inverse-$\chi^2$ prior (with hyperparameters $\nu_0=2$, and $\tau_0=1$) on the expansion coefficients yields a closed form expression for $\prob(\bm{\sigma}|\Mhi,I) $. We thus have
\begin{equation}
  p(\Mhi|\bm{\sigma},I) \propto p(\Mhi|I) \prod_{i=1}^{K} \left( \tau_i^{\nu_i}  \prod_{n \in [1,2]} Q(k_i)^{n} \right)^{-1},
  \label{eq:posterior}
\end{equation}
where $\tau_i$ and $\nu_i$ are given by
\begin{equation}
\nu_i = \nu_0 + n_c,
\label{eq:ndof_learn}
\end{equation}
and
\begin{equation}
\nu_i\tau_i^2 = \nu_0\tau_0^2 + c_1(k_i)^2+c_2(k_i)^2.
\label{eq:scale_learn}
\end{equation}
and $n_c=2$ is the number of order-by-order differences $c_1(k_i)$ and $c_2(k_i)$ for each of the $K$ momenta. The last step before we can evaluate the posterior for the breakdown scale is to express our prior $p(\Mhi|I)$. To begin with, we adopt scale-invariant log-uniform distribution across a rather large interval of possible values $\Mhi \in (m_{\pi}/40,40m_{\pi})$. 
\paragraph*{\bf Results --}
We find posteriors at NLO and \NNLO for the breakdown scale $\Mhi$ as shown in Fig.~\ref{fig:EFT_breakdown}. The \NNLO posterior is slightly more precise, as expected. Indeed, the NLO posterior is conditioned on less data as the inverse-product over the orders in Eq.~\ref{eq:posterior} is truncated at $n=1$, which also modifies Eqs.~\ref{eq:ndof_learn}-\ref{eq:scale_learn} by $n_c=1$ and $c_2(k_i)=0$ for all $k_i$.

The inferred breakdown scale is consistent with the canonical scale separation that pionless EFT is predicated on. We find median values for $\Mhi$ at 1.1$\Mpi$ and 1.4$\Mpi$ at NLO and \NNLO, respectively. Moreover, the order-by-order estimates of $\Mhi$ are consistent with each other as the NLO and \NNLO posteriors overlap within the 68\% degree of belief (dob) intervals. This is the main result of our work.
\begin{figure}[t]
  \centerline{\includegraphics[width=0.9\columnwidth,angle=0,clip=true]{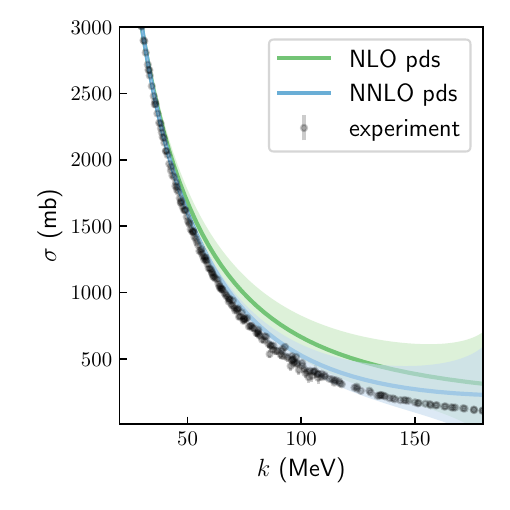}}
  \caption{\label{fig:EFT_prediction_NLO_NNLO_error} Pionless EFT predictions for the total cross section at NLO and \NNLO including 68\% DoB intervals. Loop diagrams are regularized using pds, the coupling constants are calibrated to reproduce the effective range expansion, and we employ $\Mhi=1.4\Mpi$. Experimental cross section data is from the Granada database~\cite{NavarroPerez:2013mvd,NavarroPerez:2013usk}.}
\end{figure}

Our results are largely robust with respect to physically motivated variations of our prior assumptions. Modifying the prior for the EFT expansion parameters such that $\mathbb{P}(\bar{c}^2 \leq 1) = 0.62$ by setting $\nu_0=1$ and $\tau_0=1/2$, the posterior for $\Mhi$ is also shifted to somewhat lower values, and at \NNLO we find a 68\% DoB interval $[80,179]$ MeV and median $\Mhi =\Mpi$. The NLO distribution sits at slightly lower values, as in the previous case. Regarding the soft-scale function $f(k;r)$ in Eq.~\ref{eq:soft_scale_function}, for $r>6$ the resulting inference barely changes at all, as expected from the functional form of $f(k;r)$. However, for $r=1$ we find median values of the breakdown scale at $\Mpi$ for both NLO and \NNLO. Using a a uniform distribution for the prior $p(\Mhi|I)$ across the interval $[\Mpi/2,2\Mpi]$, we find median values $\Mhi = 1.5 \Mpi$ at both NLO and \NNLO, with other characteristics of the distributions remaining largely unchanged. The median values for the breakdown scale is also robust with respect to the choice of scattering momenta at which we compute cross sections. However, increasing the number of momentum values $K$ can lead to artificially narrow posteriors, informed by correlated data, which violate the iid assumption underpinning our likelihood.

\begin{figure}[t]
\centerline{\includegraphics[width=0.9\columnwidth,angle=0,clip=true]{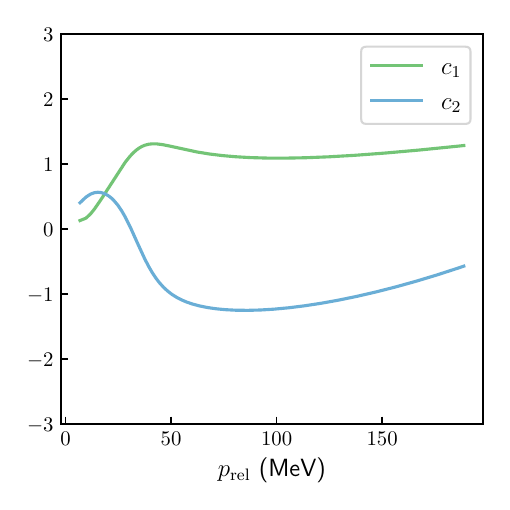}}
  \caption{\label{fig:EFT_expansion} EFT expansion coefficients $c_1$ and $c_2$ as extracted from the order-by-order predictions in Fig.~\ref{fig:EFT_prediction_NLO_NNLO_error}. }
\end{figure}

\begin{figure}[t]
\centerline{\includegraphics[width=0.9\columnwidth,angle=0,clip=true]{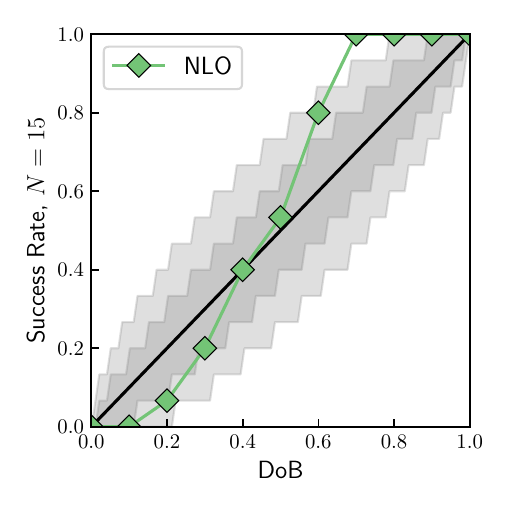}}
  \caption{\label{fig:emp_cov} Consistency plot for 15 NLO predictions at equally spaced lab scattering energies up to 75 MeV and varying DoB between 0\% and 100\%. The shaded bands represent 68\% and 95\% confidence intervals for the success rates.}
\end{figure}

We inspect the pionless EFT predictions for the total cross section and quantify the corresponding truncation error up to \NNLO, assuming $\Mhi=1.4~\Mpi$ and the prior in Eqs.~\ref{eq:prior1}-\ref{eq:prior2}. The results are shown in Fig.~\ref{fig:EFT_prediction_NLO_NNLO_error}. There is a clear trend of order-by-order improvement, and a rather good reproduction of experimental data up to $k=50(100)$ MeV for NLO(\NNLO). As a representative example, we also show the corresponding $c_1(k)$ and $c_2(k)$ EFT expansion coefficients in Fig.~\ref{fig:EFT_expansion}.
The resulting expansion coefficients are of natural size, in agreement with our expectations. The resulting EFT truncation error is also reasonable, and we quantify this using a consistency plot, see Fig.~\ref{fig:emp_cov}. The procedure for computing a consistency plot of this kind is outlined in detail in Ref.~\cite{Melendez:2017phj}. In brief, we compare the coverage of the NLO truncation error with the \NNLO prediction, and we do so at 15 equally spaced lab scattering energies up to 75 MeV. The resulting NLO coverage is within the sampling error, assumed to follow a binomial distribution. For lower (higher) DoBs the NLO truncation error can be considered too small (large). This result is a further indication of a self-consistent and robust statistical analysis.

We carried out the same analysis for pionless EFT with cutoff regularization. We find that the inferred breakdown scale of 1.4$\Mpi$ at NNLO is robust for regularization cutoff values $\gtrsim 1.5$~fm${}^{-1}$, while the NLO results show somewhat stronger variation with the cutoff. For lower values of the cutoff, the median value of the breakdown scale increases.
\paragraph*{\bf Summary -}
In this work, we inferred a median value $\Mhi \approx 1.4~\Mpi$ for the breakdown scale of pionless EFT in pds regularization. Our statistical analysis is conditioned on order-by-order predictions for the total cross section up to \NNLO. Our analysis is robust with respect to variations of our assumptions: natural expansion coefficients, iid cross section predictions, and a log-uniform prior for $\Mhi$. This result agrees with the canonical assumption that the pion sets the breakdown scale for an EFT that integrates out this mass scale in the nuclear interaction. However, it can also be considered somewhat large considering the pion cut at $\Mpi/2$ explicitly present in the $S$-wave projected one-pion exchange potential, see, e.g., Ref.~\cite{Mathelitsch:1984hq} for one of many works relating the effective range expansion and its convergence radius to the pion cut. Therefore, further study of the breakdown scale at higher orders of the pionless EFT is warranted but will come at the cost of additional LECs. For instance, at third order---one order beyond what was considered here---the $S$-wave shape parameter and the $P$-wave scattering length enter the calculation of the total cross section. An operator leading to $S$ to $D$ wave mixing enters the calculation of the total cross section at the fourth order. However, this would also enable meaningful predictions of spin-polarized cross sections. Extension to pionless EFT analyses of neutron-deuteron scattering, where the renormalization of 3-nucleon observables up to \NNLO is well understood and experimental data is abundant, should be straightforward.

Future work should address the inference sensitivity to calibrating the low-energy constants to cross-section data instead of the effective range parameters. The effective range parameters depend on the choice for the expansion of $k \cot \delta$ shown Eqs.~\eqref{eq:pole_expansion} and \eqref{eq:threshold_expansion}. Direct calibration of the LECs to experimental data might deviate from these expansions and potentially influence the order-by-order improvement of EFT predictions. It would also be interesting to marginalize over such parametric uncertainties, as well as the uncertainty in $\Mhi$~\cite{Bub:2024gyz}.

As these developments in Bayesian parameter estimation and uncertainty quantification continue, they will not only enhance uncertainty quantification of pionless EFT calculations but also facilitate Bayesian model mixing of pionless EFT and pionful EFT~\cite{Epelbaum:2008ga,Machleidt:2011zz,Hammer:2022lhx}. Mixture EFTs, combining the strengths of both EFTs, might improve the inferences conditioned on these respective EFTs. This could, e.g., prove useful in cases like proton-proton fusion, where a unified and improved uncertainty estimate for the the driving reaction rate in the Sun is desirable~\cite{Acharya:2024lke}. Finally, we remark that the approach we have used here can also be extended to the other areas where short-range EFT has been applied, provided some experimental data is available for model calibration. For example, inferring the breakdown scale of halo EFT would provide new insights into the physics of weakly bound nuclear systems.

\begin{acknowledgments}
In this work, we used the Python package 'gsum'~\cite{Melendez:2019izc} to generate most of the plots. This work was supported by the Swedish Research Council (Grants No.~2020-05127), the National Science Foundation (Grant Nos. PHY-2111426 and PHY-2412612), the Office of Nuclear Physics, and the  US Department of Energy (Contract No. DE-AC05-00OR22725).
\end{acknowledgments}

\bibliography{bibliography} 

\end{document}


\title{Supplemental Material: Quantifying the breakdown scale of pionless effective field theory}

\author{Andreas Ekström}
\affiliation{Department of Physics, Chalmers University of Technology, SE-412 96, Göteborg, Sweden}
\author{Lucas Platter}
\affiliation{Department of Physics and Astronomy, University of Tennessee, Knoxville, Tennessee 37996, USA}
\affiliation{Physics Division, Oak Ridge National Laboratory, Oak Ridge, Tennessee 37831, USA}

\begin{abstract}
This supplemental material briefly discusses the renormalization of the pionless effective field theory low-energy constants employed in the main text. As renormalization of pionless effective field theory with dimensional regularization has been extensively discussed elsewhere, we concentrate on the renormalization with a sharp momentum space cutoff. 
\end{abstract}

\maketitle
\section{Pionless effective field theory}
For completeness, we give again the Lagrangian for the pionless effective field theory (EFT). The free theory is described by the single-nucleon Lagrangian
\begin{align}
  \label{eq:L0}
  \mathcal{L} = N^\dagger ( i\partial_t +\frac{\nabla^2}{2m}) N
  +\sum_{\alpha = s,t}\mathcal{L}_\alpha +\mathcal{L}^{sd} \ldots~,
\end{align}
where $N$ denotes the (spin- and isospin spinor) nucleon field, $\alpha$ denotes either the spin-singlet channel, s, or the sprin-triplet channel, t, and $m$
is the nucleon mass.  The interaction part is written as 
\begin{align}
  \label{eq:Lagrangian}
  \nonumber
  \mathcal{L}_\alpha &=
          - C^\alpha_0 (N^T P_i^\alpha N)^\dagger (N^T P_i^\alpha N)\\
 &         +\frac{C^\alpha_2}{8}\left[\left(N^T P_i^\alpha N \right)^\dagger
    \left(N^T\overleftrightarrow{\nabla}^2 P_i^\alpha N\right) +\rm{h.c.}\right]
  \ldots
\end{align}
where $P_i^s= \sigma_2 \tau_2 \tau_i/\sqrt{8}$ ($P_i^t= \tau_2 \sigma_2 \sigma_i/\sqrt{8}$)  is the singlet (triplet)
channel projector. The operators shown above are sufficient to carry
calculations up to next-to-next-to-leading order (N2LO).

For perturbative calculations, we follow Ref.~\cite{Chen:1999tn} and expand the low-energy constants in
powers of the expansion parameter $Q$
\begin{align}
  \label{eq:lec-expansion}
  C_0 &= C_{0,-1} + C_{0,0} + C_{0,1}+\ldots~,\\
  C_2 &= C_{2,-2} + C_{2,-1} + \ldots~,\\
  C_4 &= C_{4,-3} + \ldots~.
\end{align}
\section{Calculation of amplitudes}
The LO scattering amplitude $\mathcal{A}_0$ is obtained by summing up all possible two-nucleon
scattering diagrams that contain the leading vertex in a given channel. It is related to the amplitude $t$ used in Eq.~(5) of the main via  $t = \frac{m}{4\pi}\mathcal{A}$. We write this infinite sum for the amplitude $\mathcal{A}$ as
\begin{align}
\label{eq:ALO}
  i \cA_0(E) &= -i C_{0,-1}\sum_{n=0}^\infty (-i C_{0,-1} i I_0)^n 
  \frac{i}{-\frac{1}{C_{0,-1}}+ I_0}~,
\end{align}
where $iI_0(E)$ denotes two-nucleon loop diagram in the center-of-mass system
with energy $E=p^2/m$, $p$ being the relative momentum in the incoming channel.
\begin{equation}
  \label{eq:LO-loop}
  i I_0 =\int \frac{d^4 q}{(2\pi)^4} i S(E+q_0, {\bf q}) i S(-q_0, -{\bf q})~.
\end{equation}
After performing the contour integration to solve the $q_0$ part, We
solve the remainder of this integral with a momentum space cutoff
$\Lambda$. We obtain
\begin{align}
  \label{eq:LO-Loop-2}
  \nonumber
  I_0 &= m  \int \frac{d^3 q}{(2\pi)^3}\frac{1}{p^2 - q^2 +i \epsilon}\\
  &= -\frac{m}{2\pi^2} (\Lambda(1 -\frac{p^2}{\Lambda^2} - \frac{p^4}{3\Lambda^4})+ \frac{i \pi p}{2})
\end{align}
where we have included the first $1/\Lambda$ corrections that we need
to account for when calculating the amplitude up to N2LO. We can insert this into
Eq.~\eqref{eq:ALO} to get
\begin{align}
  \nonumber
  i \cA_0 &=
  \frac{4\pi}{m}\frac{i}{-\frac{4\pi}{m C_0 }-\frac{2}{\pi}\Lambda +\frac{2 p^2}{\pi\Lambda}+\frac{2 p^4}{3 \pi \Lambda^3}-ip}~.
\end{align}
We expand in powers of $1/\Lambda$ and obtain the LO amplitude
\begin{multline}
  \label{eq:ALO-wo-loop}
  \cA_0 = \frac{4\pi}{m}\frac{1}{-\frac{4\pi}{m C_{0,-1} }-\frac{2}{\pi}\Lambda-ip}\\
  \times\left[ 1
    -\frac{2 p^2}{\pi \Lambda}\left(\frac{1}{-\frac{4\pi}{m C_{0,-1} }-\frac{2}{\pi}\Lambda-ip}\right)^2\right.\\
\left.+ \frac{4 p^4}{\pi^2 \Lambda^2}\left(\frac{1}{-\frac{4\pi}{m C_{0,-1} }-\frac{2}{\pi}\Lambda-ip}\right)^3+\ldots
  \right]~.
\end{multline}
We define
\begin{align}
  \label{eq:ALO-wo-loop-2}
  \cA_0^{(0)} &=\frac{4\pi}{m}\frac{1}{-\frac{4\pi}{m C_{0,-1} }-\frac{2}{\pi}\Lambda-ip}~,\\
  \cA_0^{(1)} &= -\frac{4\pi}{m}\left(\frac{1}{-\frac{4\pi}{m C_{0,-1} }-\frac{2}{\pi}\Lambda-ip}\right)^2 \frac{2 p^2}{\pi \Lambda}~,
              ~\\
\cA_0^{(2)} &= \frac{4\pi}{m}\left(\frac{1}{-\frac{4\pi}{m C_{0,-1} }-\frac{2}{\pi}\Lambda-ip}\right)^3 \frac{4 p^4}{\pi^2 \Lambda^2}~.
\end{align}
The second and third expressions must be taken into account during the renormalization process at NLO and NNLO, respectively.
We, therefore, see that a finite cutoff leads to terms that mimic finite effective range parameters. 
\begin{equation}
  p \cot \delta =-\frac{4\pi}{m C_{0,-1} }-\frac{2}{\pi}\Lambda~,
\end{equation}
at LO with no momentum-dependence of $p \cot \delta$. 

Now, suppose we renormalize to the scattering length so we get
\begin{equation}
  -\frac{1}{a}=-\frac{4\pi}{m C_{0,-1} }-\frac{2}{\pi}\Lambda~,
\end{equation}
which gives
\begin{equation}
  C_{0,-1} = \frac{4\pi}{m}\left(\frac{1}{a}-\frac{2}{\pi}\Lambda\right)^{-1}~.
\end{equation}
\section{Loop Integrals}
The calculation of the amplitude at different orders requires the calculation of loop integrals. The result for the simplest loop integral has already been given in Eq.~\eqref{eq:LO-Loop-2}. At higher orders, we encounter loop diagrams of the form
\begin{align}
\label{eq:loop_general}
    I_{2n} = m\int \frac{d^3 q}{(2\pi)^3}\frac{q^{2n}}{p^2 -q^2+i\epsilon}~.
\end{align}
It is easy to show that such loop integrals can be evaluated by using the formula
\begin{align}
    I_{2n} = -\frac{m}{2\pi^2}\frac{1}{2n+1}\Lambda^{2n+1} + p^2 I_{2n-2}~.
\end{align}
It is, therefore convenient to define
\begin{align}
L_{2n+1}=\frac{m}{2\pi^2}\frac{1}{2n+1}\Lambda^{2n+1}~.
\end{align}
\section{Renormalized expressions}
\subsection{Next-to-leading order}
Cutoff regularization and renormalization at next-to-leading order (NLO) was discussed in \cite{Emmons:2016myr}. At NLO, perturbative insertions of the subleading operators leads to the amplitude
\begin{multline}
  \label{eq:Anlo_sum}
  i \cA_1 =\left( \frac{\cA_0}{C_{0,-1}}\right)^2\Bigl[-i \frac{m p^2}{2\pi^2\Lambda}{C_{0,-1}}^2\\
           -i C_{2,-2} \left(p^2
                  - L_3C_{0,-1}
                   \right)
            - i C_{0,0} \Bigr]~,
\end{multline}
The first term in the above expression is the residual cutoff dependence from Eq.~\eqref{eq:ALO-wo-loop} contributing at this order.

Matching to the threshold expansion as we do in the singlet channel gives leads to the two equations
\begin{align}
  \label{eq:nlo_matching}
  \nonumber
 C_{2,-2} &=-(C_{0,-1})^2\left(\frac{m}{2 \pi^2 \Lambda}-\frac{m}{4\pi}\frac{r_s}{2}\right)~,\\ 
  C_{0,0} & = C_{2,-2} C_{0,-1}L_3 
\end{align}
In the spin-triplet channel, we can use the expansion around the pole (see Eq.~(3) of main text) and get
\begin{align}
  \label{eq:nlo_matching_triplet}
C_{2,-2}=&-C_{0,-1}^2\left(\frac{m}{2\pi^2\Lambda}-\frac{m}{4\pi}\frac{\rho_t}{2} \right)~,\\
 C_{0,0}&=C_{2,-2} C_{0,-1} L^3+{C_{0,-1}}^2\frac{m}{4\pi}\frac{1}{2}\gamma^2\rho_t~.
\end{align}
\subsection{Next-to-next-to-leading order}
At NNLO, we find for the LECs in the threshold expansion that we use in the singlet channel
\begin{align}
  \label{eq:lecs-nnlo-cutoff-threshold}
  C_{4,-3} &= \frac{C_{2,-2}^2}{C_{0,-1}}~,\\
  C_{2,-1} & = \frac{11 C_{2,-1}^2 L_3}{4}~,\\
  C_{0,1} & = \frac{1}{4} C_{2,-2}^2 \left( 8 C_{0,-1} L_3^2 +L_5\right)
\end{align}
When we match to the pole expansion that we use in the triplet channel, we find
\begin{align}
  \label{eq:lecs-nnlo-cutoff-pole}
  C_{4,-3} &= \frac{C_{2,-2}^2}{C_{0,-1}}~,\\
  C_{2,-1} &= \frac{1}{4\pi}\left(11 C_{2,-2}^2 L_3 \pi + m \rho \gamma^2 C_{0,-1} C_{2,-2} \right)~,\\
  \nonumber
  C_{0,1}  &= \frac{1}{64 \pi^2}\left(
             128 C_{0,-1} C_{2,-2}^2 L_3^2 \pi^2 +16 C_{2,-2}^2 L_5 \pi^2\right.\\
           &\qquad             \left.+24 C_{0,-1}^2 C_{2,-2} L_3 m \pi \rho\gamma^2 +C_{0,-1}^3 m^2 \rho^2 \gamma^4
             \right)
\end{align}
\bibliography{bibliography}